\documentclass[aps,pra,showpacs,floatfix,superscriptaddress,nofootinbib,longbibliography,11pt]{revtex4-1}
\usepackage{xr}
\usepackage{amsmath,amssymb,physics}
\usepackage[toc,page]{appendix}
\usepackage{verbatim,graphicx,mathtools}
\usepackage{txfonts}
\usepackage{color}
\usepackage[all]{xy}
\usepackage{subfigure}
\usepackage{xcolor}
\usepackage{hyperref}
 \usepackage[bottom]{footmisc} 

\usepackage{slashed}
\usepackage{mathtools}
\usepackage{float}
\usepackage{empheq}
\usepackage{bm}
\usepackage{framed}
\usepackage{comment}
\usepackage{xcolor}

\def\pg{\paragraph}

\def\ie{\emph{i.e.} }

\def\Z{\mathbb{Z}}

\def\cE{\mathcal{E}}

\def\cO{\mathcal{O}}
\def\cP {\mathcal{P}}

\def\p{\partial}

\def\/{\over}
\def\ov{\over}

\def\rn{\rangle}
\def\ln{\langle}

\def\t{\theta}

\def\e{\epsilon}

\def\b{\beta}
\def\d{\delta}
\def\k{\kappa}
\def\g {\gamma}

\def\mn{{\mu\nu}}

\def\n {\nabla}

\def\D{\Delta}

\def\ra{\rightarrow}

\def\Tr{\mathrm{Tr}}

\def\r{\mathrm}

\def\_{\hspace{2cm}}
\def\-{\\\notag}
\def\={&=&}

\newcommand{\bea}{\begin{eqnarray}}
\newcommand{\eea}{\end{eqnarray}}

\newcommand{\bpm}{\begin{pmatrix}}
\newcommand{\epm}{\end{pmatrix}}

\newcommand{\bit}{\begin{itemize}}
\newcommand{\eit}{\end{itemize}}

\newcommand{\ben}{\begin{enumerate}}
\newcommand{\een}{\end{enumerate}}

\newcommand\bsp{\begin{split}}
\newcommand\esp{\end{split}}

\def\le{\left}
\def\ri{\right}

\def\qq{\qquad}

\def\cos{\r{cos}}

\newcommand{\be}{\begin{eqnarray}}
\newcommand{\ee}{\end{eqnarray}}

\newcommand{\bmat}{\left ( \begin{array}{cc} }
	\newcommand{\emat}{\end{array} \right ) }

\def\Tr{\textrm{Tr}}

\newcounter{jvct}

\newcommand{\beq}{\begin{equation}}
\newcommand{\beqs}{\begin{equation*}}
\newcommand{\eeq}{\end{equation}}
\newcommand{\eeqs}{\end{equation*}}

\allowdisplaybreaks

\usepackage{scalerel,stackengine}

\newcommand\reallywidehat[1]{%
\savestack{\tmpbox}{\stretchto{%
  \scaleto{%
    \scalerel*[\widthof{\ensuremath{#1}}]{\kern-.6pt\bigwedge\kern-.6pt}%
    {\rule[-\textheight/2]{1ex}{\textheight}}
  }{\textheight}%
}{0.5ex}}%
\stackon[1pt]{#1}{\tmpbox}%
}

\begin{document}

\begin{center}
\vspace{2cm}
\end{center}

\title{Euclidean wormhole in the SYK model}
\author{Antonio M. Garc\'\i a-Garc\'\i a}
\email{amgg@sjtu.edu.cn}
\affiliation{Shanghai Center for Complex Physics, 
	School of Physics and Astronomy, Shanghai Jiao Tong
	University, Shanghai 200240, China}

\author{Victor Godet}
\email{victor.godet.h@gmail.com}
\affiliation{Institute for Theoretical Physics, University of Amsterdam, 1090 GL Amsterdam, Netherlands}

\begin{abstract} 

\vspace{3.5cm}
We study a two-site Sachdev-Ye-Kitaev (SYK) model with complex couplings, and identify a low temperature transition to a gapped phase characterized by a constant in temperature free energy. This transition is observed without introducing a coupling between the two sites, and only appears after ensemble average over the complex couplings. We propose a gravity interpretation of these results by constructing an explicit solution of Jackiw-Teitelboim (JT) gravity with matter: a two-dimensional Euclidean wormhole whose geometry is the double trumpet. This solution is sustained by imaginary sources for a marginal operator, without the need of a coupling between the two boundaries. As the temperature is decreased, there is a transition from a disconnected phase with two black holes to the connected wormhole phase, in qualitative agreement with the SYK observation. The expectation value of the marginal operator is an order parameter for this transition. This  
illustrates in a concrete setup how a Euclidean wormhole can arise from an average over field theory couplings.

\end{abstract}\maketitle

\newpage

\section{{Introduction}}
Wormholes are geometric shortcuts that connect distant points in spacetime. Their role in the Euclidean path integral has been hotly  debated in the literature \cite{Lavrelashvili:1987jg, Hawking:1987mz, Giddings:1987cg, giddins1988, GIDDINGS1989481, coleman1988,maldacena2004,arkanihamed2007, Hertog:2018kbz,Hebecker:2018ofv}. In the context of holography, an important puzzle is that geometries connecting two boundaries indicate that the partition function of the dual field theory does not factorize \cite{maldacena2004}.  To address this problem, one could simply decide not to include these configurations. However, it has been recently realized that it is only after including Euclidean wormholes that one gets results consistent with the interpretation of black holes as ordinary quantum systems. This has been shown explicitly in Jackiw-Teitelboim gravity \cite{jackiw1985,teitelboim1983}, a two-dimensional theory of gravity capturing the low-energy dynamics of near-extreme black holes \cite{jensen2016,maldacena2016a,engels2016}, for the spectral form factor \cite{Saad:2018bqo, Saad:2019lba}, for correlations functions \cite{Saad:2019pqd}, and for the fine-grained entropy of evaporating black holes \cite{Penington:2019npb, Almheiri:2019psf, penington2020,almheiri2020, almheiri2020a}. The factorization puzzle suggests that the gravitational path integral requires some form of ensemble averaging, whose origin remains mysterious. We refer to \cite{Blommaert:2019wfy, Betzios:2019rds, marolf2020, Pollack:2020gfa, vanraamsdonk2020, Betzios:2020nry, turiaci2020,Blommaert:2020seb,Anous:2020lka,Chen:2020tes,Eberhardt:2020bgq,Stanford:2020wkf,McNamara:2020uza,alishahiha2020,Cotler:2020lxj} for recent discussions on this issue.  

This problem doesn't arise in Lorentzian signature because non-traversable wormholes have horizons, which make them consistent with the factorization of the field theory dual. In this case, wormholes are interpreted as coming from  quantum entanglement \cite{maldacena2003, Maldacena:2013xja}. It has also been recently shown that these wormholes can be rendered traversable by introducing a double trace coupling between the boundaries \cite{gao2016,maldacena2017}. In particular, Maldacena and Qi \cite{maldacena2018}, see also \cite{bak2018,kim2019}, have described an eternal traversable wormhole solution in JT gravity with a double trace deformation, and argued that a dual picture consists in two copies of a Sachdev-Ye-Kitaev (SYK) model \cite{french1970,bohigas1971,sachdev1993,kitaev2015,maldacena2016,jensen2016,jevicki2016,garcia2016,garcia2017,bagrets2016,bagrets2017} weakly coupled by a one-body operator. The system undergoes a first order transition at finite temperature from the wormhole to a two black holes phase which can also be characterized by spectral statistics \cite{garcia2019}.

In this paper, we propose a Euclidean version of this story which doesn't involve an explicit coupling between the boundaries. In section II, we show that the free energy of a two-site SYK model with complex couplings, obtained by exact diagonalization of the Hamiltonian, undergoes a low temperature transition to a gapped phase similar to that of the wormhole of \cite{maldacena2018}, and which arises only after an ensemble average over couplings. In section III, we describe a Euclidean wormhole solution of JT gravity plus matter, which does not require a coupling between the two boundaries, but is instead sustained by imaginary sources. We compute the free energy and show that this system undergoes a similar phase transition from a high temperature phase with two black holes to a low temperature wormhole phase, in qualitative agreement with the SYK behavior. We end with a discussion and conclusions in section IV.



\newpage

\section{Sachdev-Ye-Kitaev with complex couplings}

We study two uncoupled non-Hermitian SYK models with complex couplings composed by $N$ Majorana fermions in $(0 + 1)$ dimensions with infinite range interactions in Fock space. 
One of the copies is called Left ($L$) with Majoranas denoted $\psi_L$. The other copy is called Right ($R$) with Majoranas denoted $\psi_R$. Majoranas in each copy are governed by an  SYK Hamiltonian with complex couplings\footnote{Antonio M. Garc\'ia Garc\'ia thanks Zhenbin Yang for pointing out section 5.6 of Ref.~\cite{maldacena2018} where the relation between SYK models with complex couplings and Euclidean wormholes is mentioned and for suggesting the model (\ref{hami})} with the left Hamiltonian being the complex conjugate of the right one:
\begin{align}
H_{L} \, &= \, \frac{1}{4!} \sum_{i,j,k,l=1}^{N/2} (J_{ijkl}+i\kappa M_{ijkl}) \, \psi_{L,i} \, \psi_{L,j} \, \psi_{L,k} \, \psi_{L,l} \,\nonumber \\
H_{R} \, &= \, \frac{1}{4!} \sum_{i,j,k,l=1}^{N/2} (J_{ijkl} -i\kappa M_{ijkl}) \, \psi_{R,i} \, \psi_{R,j} \, \psi_{R,k} \, \psi_{R,l}
\end{align}
where $\kappa$ is a real positive number,
$\{ \psi_{A,i}, \psi_{B,j} \} = \delta_{AB}\delta_{ij}$ $(A,B=L,R)$ and $J_{ijkl}, M_{ijkl}$ are Gaussian distributed random variables
with zero average and standard deviation $\sqrt{\langle J_{ijkl}^2\rangle}  = \sqrt{\langle M_{ijkl}^2\rangle}= 4\sqrt{6}/{N^{3/2}}$, see \cite{kitaev2015,maldacena2016}. We note that both Hamiltonians are non-Hermitian and there is no explicit coupling between them.

 The combined system Hamiltonian
\be \label{hami}
H = H_L + H_R
\ee
 has a complex spectrum with complex conjugation symmetry: if $E_n$ is an  eigenvalue, its complex conjugate $E_n^\ast$ is also an eigenvalue. We shall see that despite the fact that the two copies are decoupled, the combined system  after ensemble average shares many of the  properties expected from a Euclidean wormhole. 

\begin{figure}
 \centering
      	\includegraphics[width=8cm]{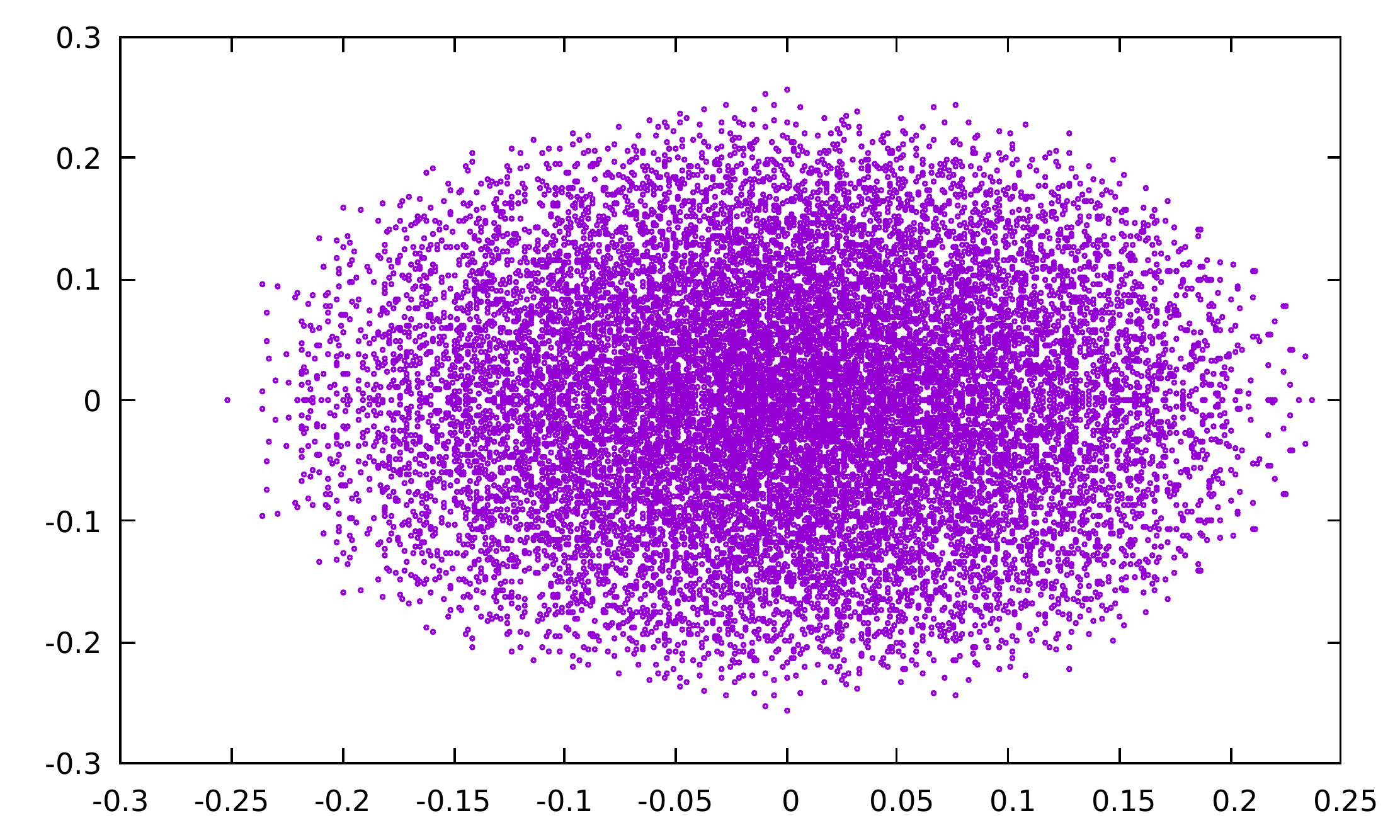}
      		\includegraphics[width=8cm]{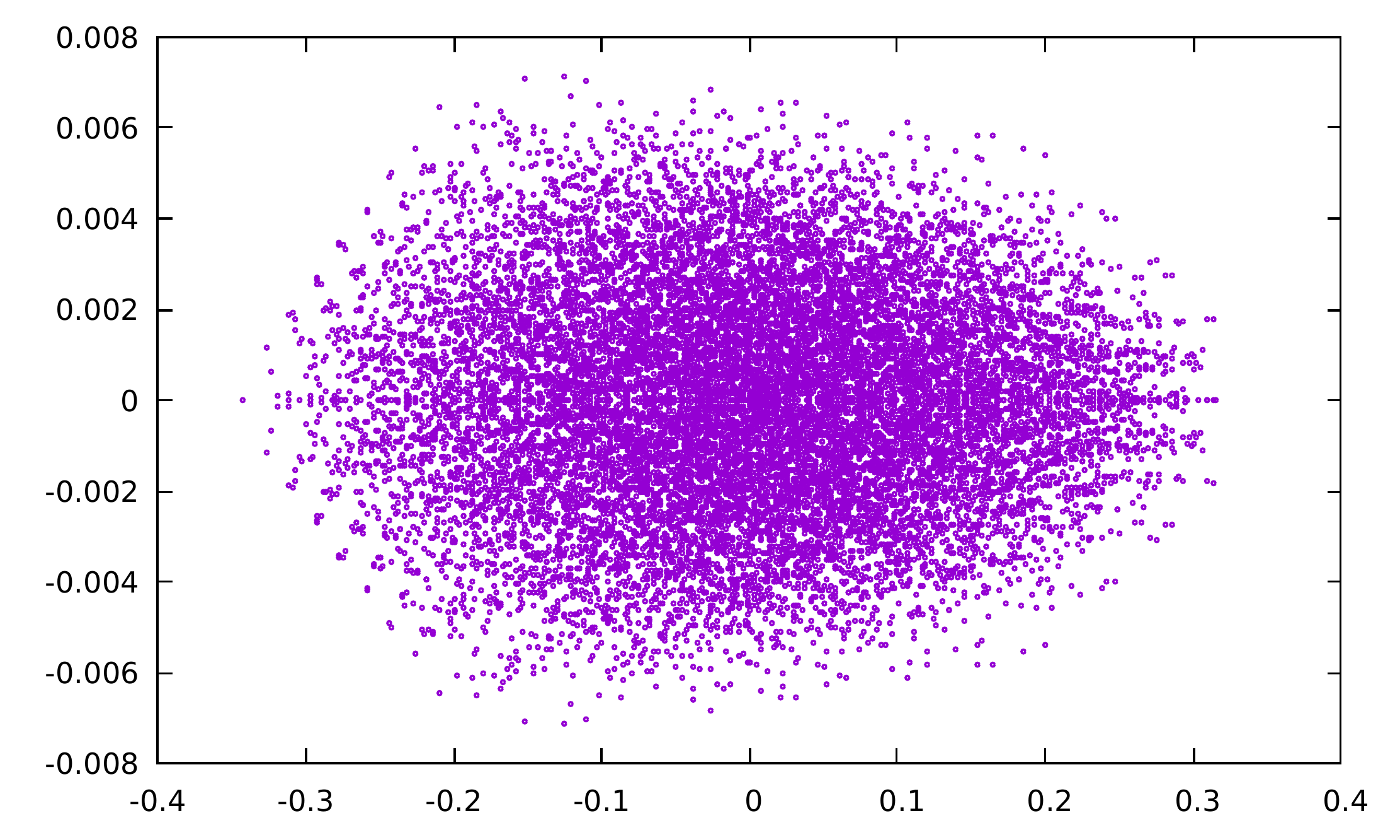}
      		\includegraphics[width=8cm]{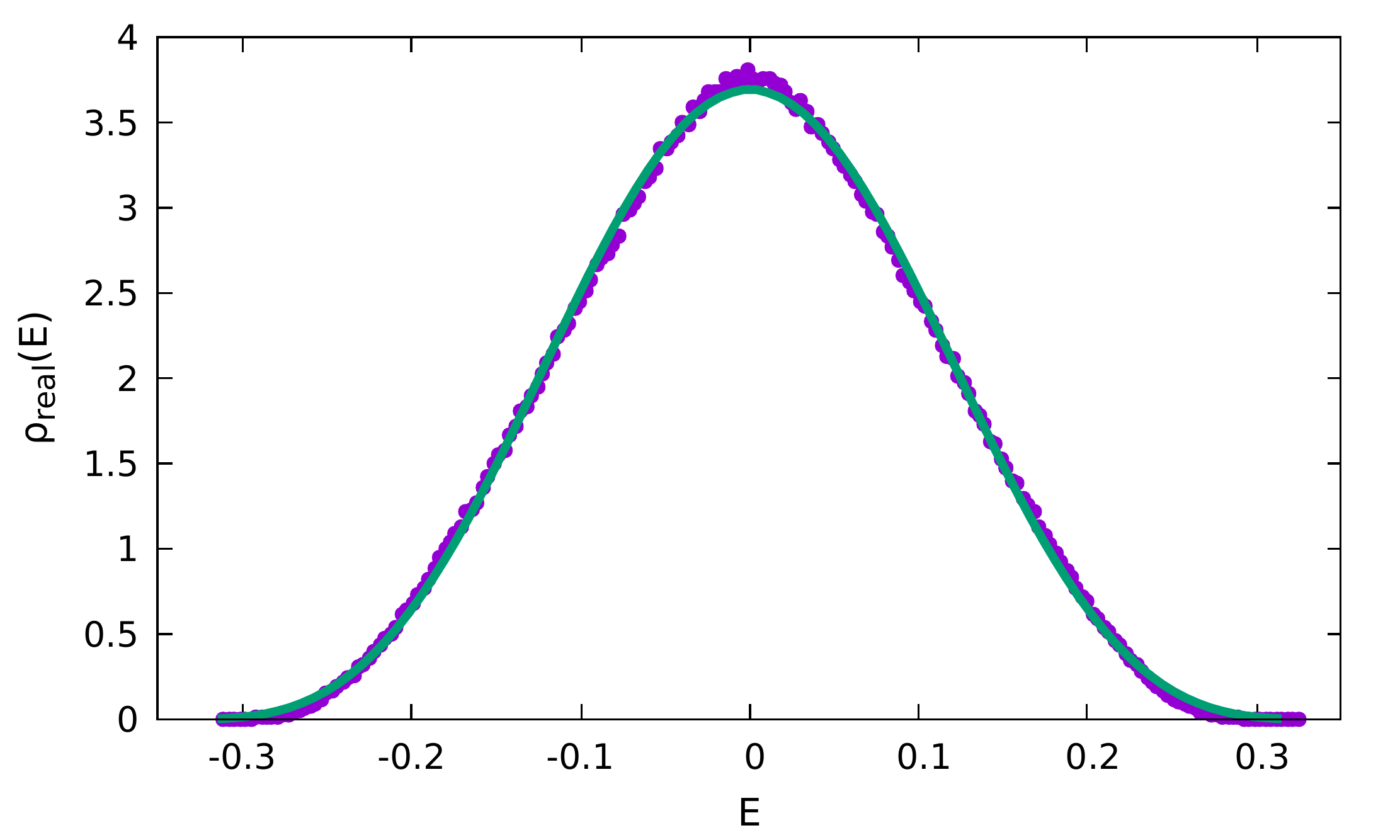}
      		\includegraphics[width=8cm]{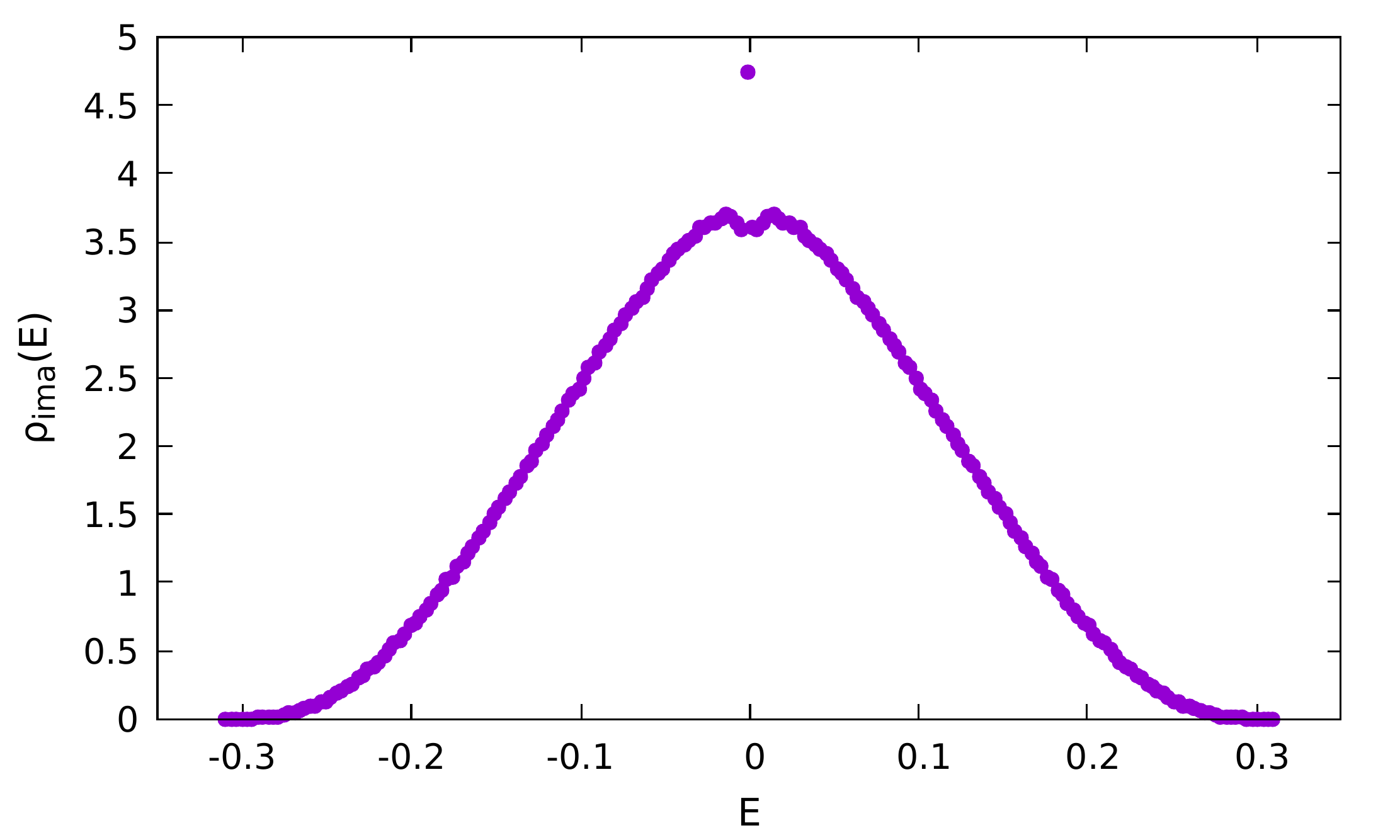}

	\caption{
		Top: Complex spectrum of the combined Hamiltonian (\ref{hami}) for $N = 30$ and, $\kappa = 1$ (left) and $\kappa = 0.1$ (right). Note the different scale of each figure.
		Bottom: Spectral density of the real (left) and imaginary (right) part of the eigenvalues for $\kappa = 1$, $N = 34$ after average over $45$ disorder realizations. The real part looks qualitatively similar to that of a single SYK model with real couplings. Indeed, it agrees well (solid line) with the analytical prediction, see $(24)$ of \cite{garcia2017}, valid everywhere except in the tail of the spectrum. The best fitting is close to the analytical prediction for   $2N$ Majoranas. Regarding the imaginary part, it is characterized by a peak at zero energy followed by a suppression for small energies whose origin at present we do not understand well.
	}
	\label{fig:spek}
\end{figure}

 We compute the spectrum by exact diagonalization techniques with $N \leq 34$. The spectrum for a single disorder realization is depicted in Fig.~\ref{fig:spek} for different values of $\kappa$ which is the only free parameter of the model. For $\kappa = 1$, the eigenvalues are distributed in the complex plane with an ellipsoid shape with axis of similar sizes. Not surprisingly, for small $\kappa = 0.1$,  although the spectrum still has an ellipsoid shape, it is mostly concentrated close to the real axis. More information is revealed, see lower plots in Fig.~\ref{fig:spek}, in the spectral density of the real and imaginary parts of the eigenvalues after $1000$ disorder realizations. The spectral density of the real part seems to be qualitatively similar to that of the SYK model with real couplings. The imaginary part shows by contrast a sharp peak at zero and a relatively small depression around zero energy. A similar peak is observed for other values of $\kappa$, though its strength, as expected, diminishes as $\kappa$ increases. We do not have a clear understanding of why some eigenvalues have zero imaginary part even in the bulk of the spectrum while there is some level repulsion around zero. The latter is likely related to conjugation symmetry of the spectrum that effectively acts as a chiral symmetry inducing level repulsion around states with zero (imaginary) energy. Real eigenvalues are not beneficial for the establishment of the wormhole phase because the absence of an imaginary part prevents the cancellations necessary for its existence. Likely, this is not important as the mentioned cancellations only occur in the infrared limit of the spectrum where the number of real eigenvalues, other than the ground state, is small. An important feature of the spectrum is that the ground state $E_0$ is always real for any $\kappa$ or strength of disorder. 

\begin{figure}
	\includegraphics[width=8cm]{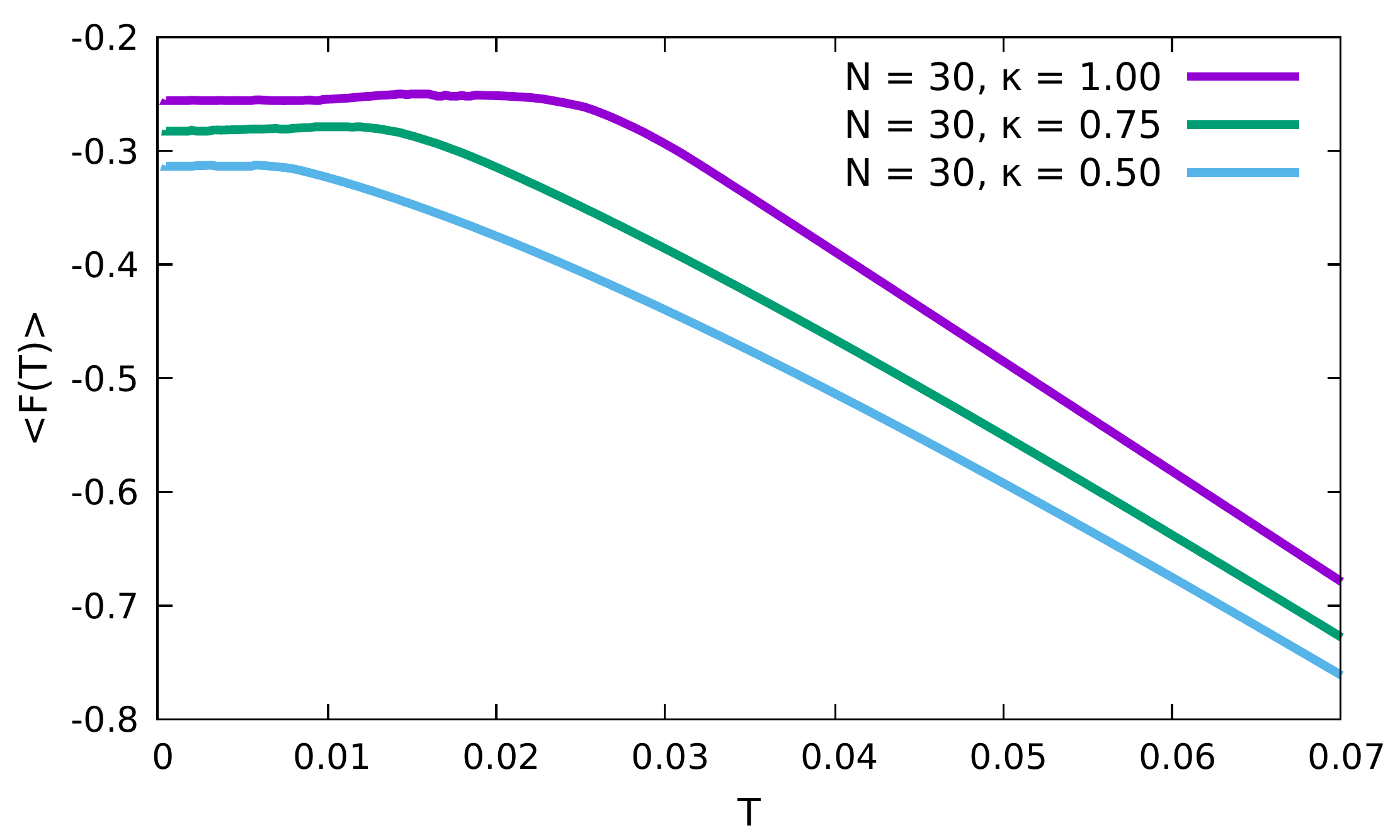}
	\includegraphics[width=8cm]{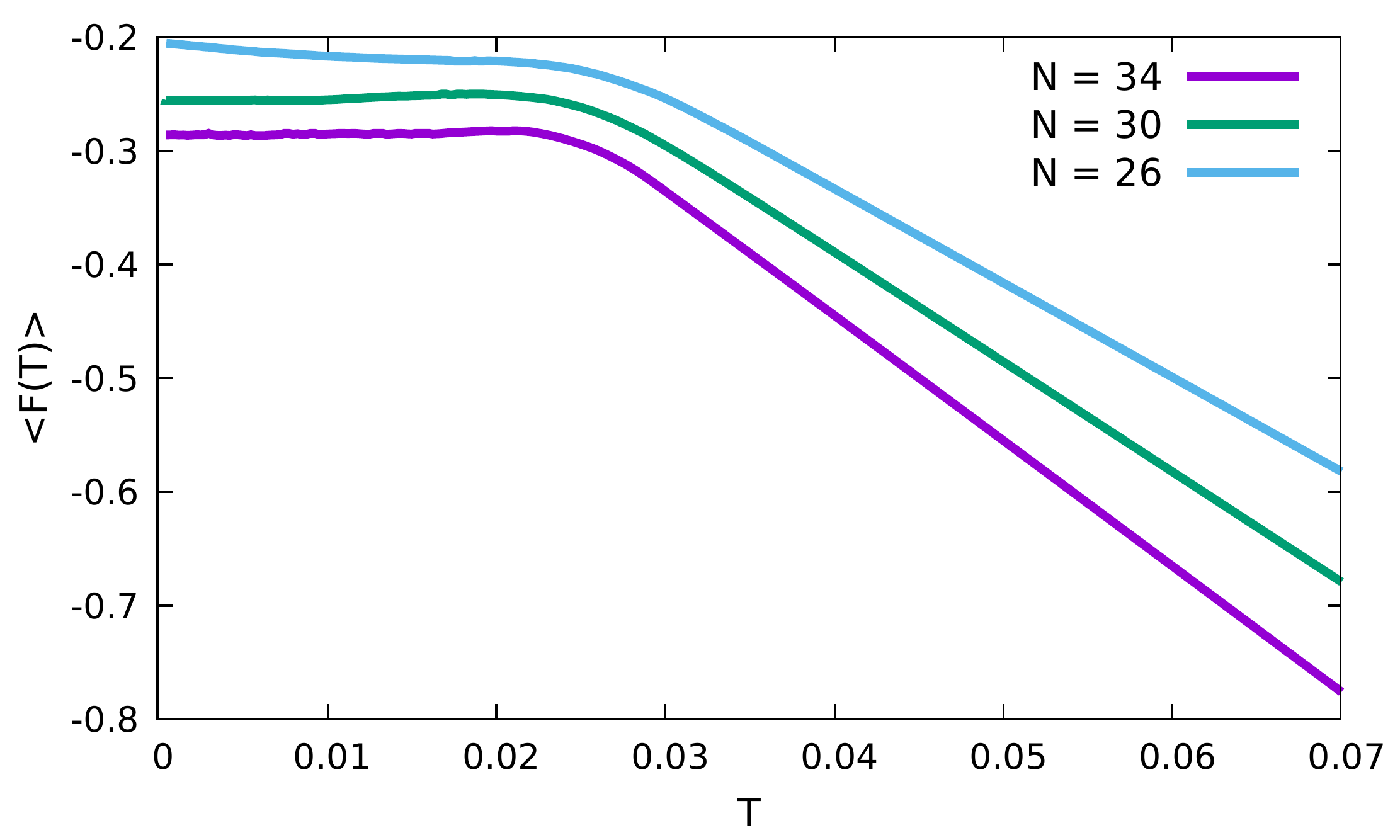}
	\caption{
		Left: Free energy after ensemble average of $300$ disorder realizations for $N = 30$ Majoranas and different strengths of the imaginary part $\kappa$.  Right: Free energy for $\kappa = 1$ and different $N$'s. 
	}
	\label{fig:freeq}
\end{figure}

We can now proceed to the calculation of the thermodynamic properties. Interestingly, because complex conjugation is a symmetry of the spectrum, the partition function of the combined system $Z(\beta)=\Tr\, e^{-\beta H}$ is real. In order to reduce statistical fluctuations, we carry out an ensemble average and compute the resulting quenched free energy:
\be 
\le\langle F(T) \ri\rangle = - T\le\langle {\log Z(\b)} \ri\rangle,
\ee
where $\b=1/T$. Results depicted in Fig.~\ref{fig:freeq} for different values of $\kappa$ show a surprising result. The free energy is constant for sufficiently low $T$ and there is a rather abrupt change at a certain critical temperature that suggests the existence of a first order transition. We note that a sharp transition only occurs in the $N \to \infty$ limit. Although numerically it is hard to reach beyond a comparatively small value of  $N \sim 34$, it is still important to assess the magnitude of finite $N$ effects in the range of available sizes. Results depicted in the right plot of Fig.~\ref{fig:freeq} show that these effects are relatively small, consistent already with  $1/N$ corrections. As is expected, these perturbative finite $N$ effects tend to increase the free energy.

Roughly speaking, the location of the kink, which corresponds to the critical temperature, seems to scale approximately as $\kappa^2$. A precise determination is difficult because for small $\kappa \ll 1$ it becomes harder to clearly see the  low temperature phase. This could be a finite $N$ effect or might signal the existence of a minimum value of $\kappa$, even in the large $N$ limit, for a gap to be formed. For large $\kappa \gg 1$, the free energy is initially constant but display  modulations (not shown) at intermediate temperatures. At present, we do not have an explanation for this intermediate phase. We can only say that it does not seem to be a statistical fluctuation that will go away in the large $N$ limit or if more disorder realizations are considered.   
 
This free energy is very similar to both the free energy of the eternal traversable wormhole studied in \cite{maldacena2018} and to that of the Euclidean wormhole described in the next section. A constant free energy signals the existence of a gap in the spectrum that separates the ground state from higher excitations of the system. The existence of the gap can be explained by the combined effect of a complex spectrum, the mentioned complex conjugation symmetry, and ensemble average as follows.

  Writing the spectrum of $H$ as, $E_{n}=a_n + ib_n$ where $a,b$ are real numbers, using the fact that if $E_n$ is an eigenvalue then $E_n^*$ is also an eigenvalue and that the ground state $E_0$ is real, we write the partition function for a given disorder realization as, $Z(\beta)=e^{-\beta E_0}+2\sum_{n}\cos(\beta b_n)\,e^{-\beta a_n}$, assuming no degeneracies. After ensemble average, the free energy becomes
  \be 
  \langle \beta F \rangle = - \left\langle {\log ( e^{-\beta E_0}+2\sum_{n}\cos(\beta b_n)\,e^{-\beta a_n})}  \right \rangle .
  \ee
  
  \begin{figure}
  	\includegraphics[width=14cm]{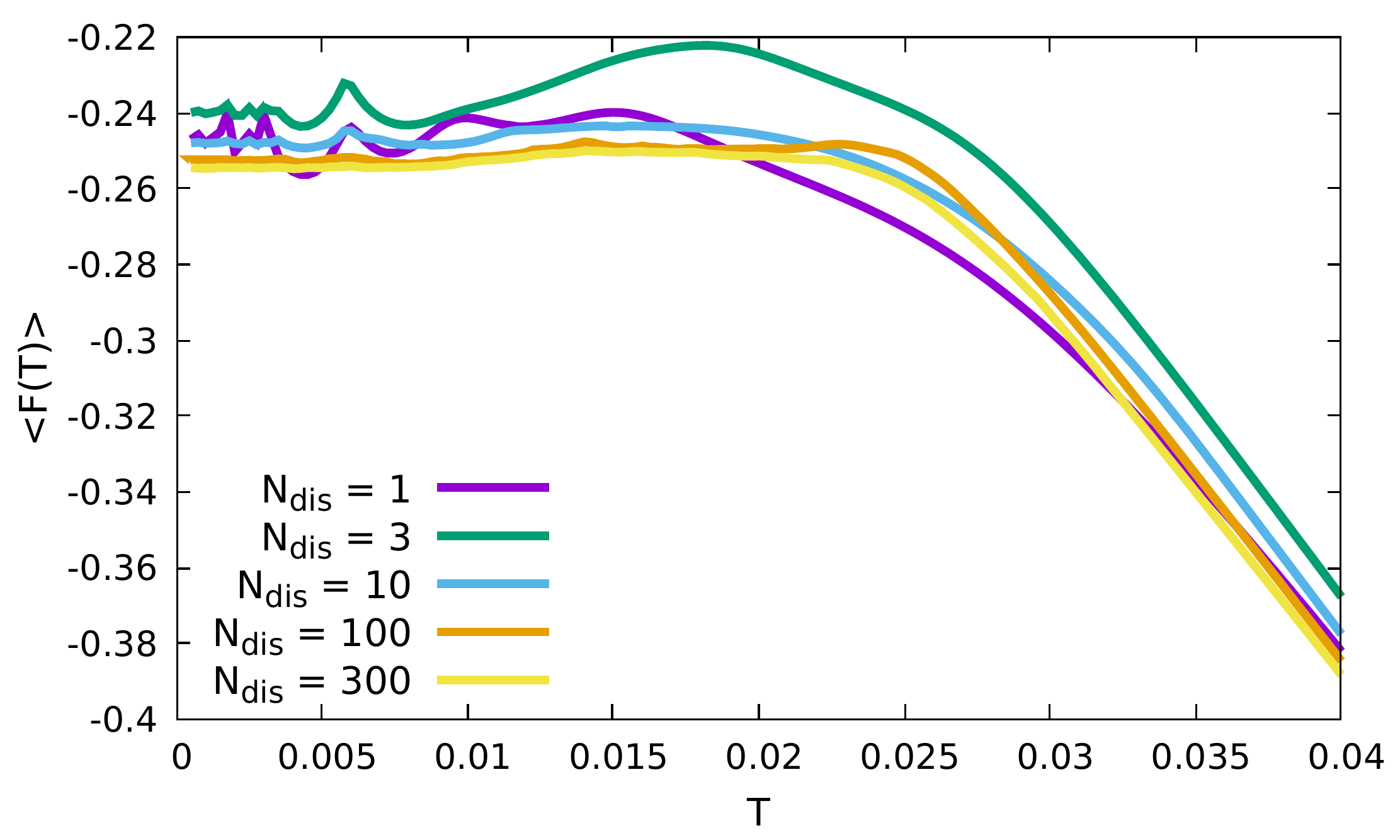}
  	\caption{
  		Dependence of free energy of the Hamiltonian  (\ref{hami}) on the number of disorder realizations $N_{dis}$ for $N = 30$ and $\kappa = 1$. A flat free energy in the low temperature limit that signals a gapped spectrum and a wormhole phase is only observed after ensemble average with a large number of disorder realizations.  
  	}
  	\label{fig:feens}
  \end{figure}
\noindent
In the high temperature limit $\beta \to 0$, the argument of the cosine function is always small and $\cos \sim 1$. As a consequence, the imaginary part of the spectrum becomes effectively irrelevant and the spectrum corresponds to that of two identical, uncoupled, systems, in this case two  SYK models with real couplings. Assuming that a gravity dual picture still applies, this region corresponds to a phase with two black holes. 

In the opposite limit $\beta \to \infty$, corresponding to low temperatures, the cosine becomes a highly oscillating function whose features depend on the form of the imaginary part of the spectrum. From the results depicted in Fig.~\ref{fig:spek}, the imaginary part of the spectrum seems to vary greatly, especially for intermediate $\kappa$, even for eigenvalues close to the ground state. In any case, the variations of the imaginary part are much faster than those of the real part. 

The effect of the ensemble average is to effectively suppress the contribution  of eigenvalues in the lower part of the spectrum, allowing the opening of a gap above the ground state $E_0$. This leads to a free energy very similar to the one recently reported for the eternal traversable wormhole \cite{maldacena2018}. The main differences is that in our setup, there is no explicit coupling between the two copies. 

\vspace{1cm}

We now address in more detail the importance of ensemble average in our results.
 In Fig.~\ref{fig:feens}, we depict results for the free energy after ensemble average with an increasing number of disorder realizations. For a small number of disorder realizations, the free energy in the low temperature limit is far from being flat. Peaks and oscillations of different frequencies are observed. Only after performing the average with a comparatively large number of realizations, the free energy becomes completely flat in this limit which is a signature of a gapped spectrum and a wormhole. It could be argued that, for a much larger number of Majoranas, which for $N \geq 36$ is numerically expensive, spectral average on a single realization of disorder is enough to flatten the free energy but at present we do not have evidence of this. Moreover, even if we could reach a much larger value of $N$, we believe that if no ensemble average is carried out, it would be necessary to perform a spectral average in a small window to smooth out fluctuations.  

In the next section, we will see that the free energy of our SYK setup is strikingly similar to that of a solution of JT gravity plus matter, a Euclidean wormhole with the geometry of the double trumpet. This led us to propose that the low-temperature phase that we observe in our SYK setup should be interpreted as a Euclidean wormhole.  
Therefore, with the present evidence, ensemble average in the field theory dual is a crucial ingredient to reproduce the expected features of a Euclidean wormhole, as expected from factorization arguments. 
Finally, let us comment on the advantage of exact diagonalization techniques with respect to the solution of Schwinger-Dyson equations in the large $N$ limit. There, the ensemble average is carried out earlier in the derivation of the equations and therefore it is not possible to determine its exact role in observing the gap at low temperature. In contrast, exact diagonalization displays  without ambiguity the central role played by the ensemble average.  

\newpage

\section{The double trumpet solution}

We have  observed that the SYK model with complex couplings has a low temperature phase which behaves like a wormhole. In this section, we propose a gravitational interpretation of this phase, as a novel Euclidean wormhole solution of JT gravity plus matter, where the crucial ingredient is the introduction of imaginary sources for a marginal operator.

The theory we consider is JT gravity with a massless scalar field, described by the action
\be
S = S_\r{JT} +S_\r{matter}~,
\ee
where we have
\bea\label{actionJT}
S_\r{JT} \= -{S_0\/2\pi}\le[{1\/2} \int d^2 x\sqrt{g}\, R + \int d\tau\sqrt{h} \,K\ri]-{1\/2} \int d^2 x\sqrt{g}\, \Phi(R+2) - \int d\tau\sqrt{h} \, \Phi (K-1)~,
\\
S_\r{matter} \= {1\/2}\int d^2 x\sqrt{g} \,(\p\chi)^2~.
\eea
JT gravity is a two-dimensional theory that captures the low-energy dynamics of higher-dimensional near-extreme black holes \cite{almheiri2015, maldacena2016a,Almheiri:2016fws, Nayak:2018qej, Castro:2019crn}. The parameter $S_0$ is interpreted as the extremal entropy and is taken to be large. In the Euclidean path integral \cite{Saad:2019lba}, the first term in \eqref{actionJT} gives the topological contribution $e^{-S_0(2g+n-2)}$ to a geometry of genus $g$ with $n $ boundaries, so a large value of $S_0$ suppresses higher genus contributions.

\begin{figure}[H]
	\centering
	\includegraphics[width=16cm]{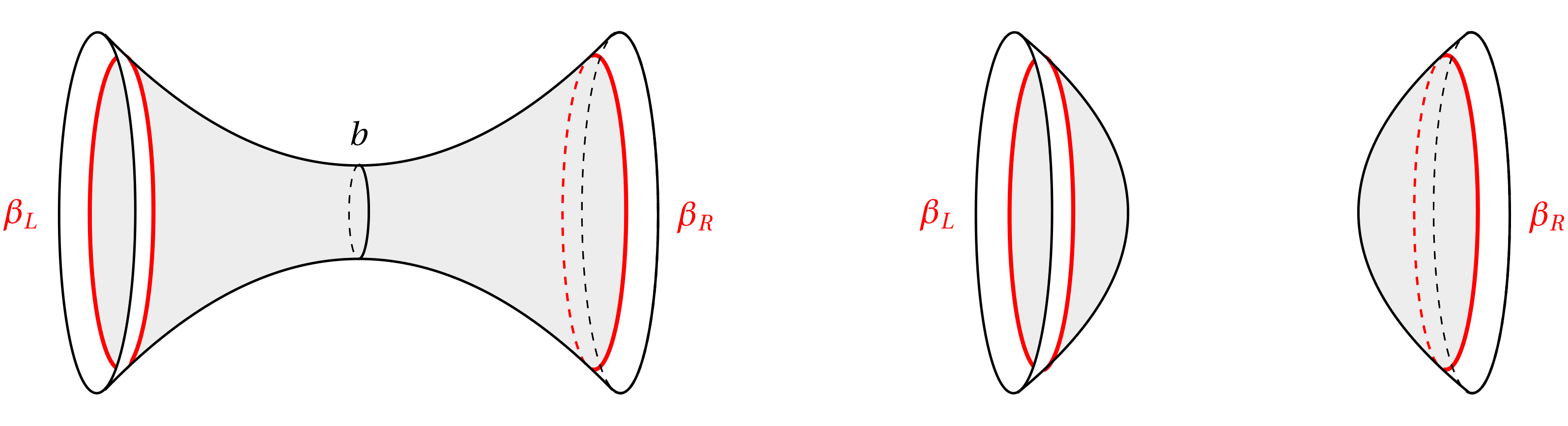}
	\caption{Left: Double trumpet geometry corresponding to the wormhole phase. Right: Hyperbolic disks corresponding to the disconnected phase with two black holes. We observe a low temperature transition between these two phases.}\label{Fig:DT}
\end{figure}

The solution we will describe has the geometry of the double trumpet: a Euclidean geometry with $R=-2$ and with two asymptotic boundaries. It can be described by the metric 
\be\label{doubleTrumpet}
ds^2= {1\/\r{cos}^2\rho} \le(d\tau^2+d\rho^2\ri)~,\qq -{\pi\/2} \leq \rho\leq {\pi\/2}~,\qq \tau\sim \tau+b~,
\ee
which is just global AdS$_2$ with periodically identified Euclidean time. The proper length of the geodesic at $\rho=0$ is called $b$ and is the only parameter of the geometry. In JT gravity, we also have a left and right inverse temperatures $\b_L$ and $\b_R$, defined as the  periods of the respective boundary times, see Fig.~\ref{Fig:DT}.

The double trumpet geometry is not an ordinary solution of JT gravity plus matter.  The crucial additional ingredient that we use here is to turn on imaginary sources for a massless scalar field. The motivation for this comes from the SYK story described in the previous section. There, a wormhole-like phase appears after adding an imaginary part to the couplings. We can interpret that as a deformation of the left and right Hamiltonian by an operator
\be
\d H_L  = -i\k  M_L,\qq \d H_R= i\k M_R~,\qq M\equiv \frac{1}{4!} \sum_{i,j,k,l=1}^{N/2} M_{ijkl} \, \psi_{i} \, \psi_{j} \, \psi_{k} \, \psi_{l} ~.
\ee
We aim to investigate the effect of a qualitatively similar deformation on the gravity side. For that purpose, we deform the JT gravity action by a scalar operator $\cO$ on each side
\be
\d S= -i k\int_{S^1_L} d\tau \,\cO_R(\tau)+i k\int_{S^1_R} d\tau \,\cO_L(\tau)~,
\ee
where the source is $i k$ on the left boundary and $- i k$ on the right boundary. Using the standard AdS/CFT dictionary, this should correspond to fixing the asymptotic value of some bulk scalar field $\chi$.   

To obtain a wormhole solution in JT with matter, it turns out that we should take $\cO$ to have conformal dimension $\D=1$ so that $\chi$ is a massless scalar field in the bulk. This is explained in Appendix \ref{app:massive}. The condition we impose is then
\be\label{chiDirichlet}
\lim_{\rho\ra {\pi/2}} \chi = i k,\qq \lim_{\rho\to-{\pi/2}} \chi =- i k~,
\ee
and we will see that this makes the wormhole solution possible. To be clear, we are not claiming that this follows from a precise holographic duality. It must be understood as a setup in Euclidean JT gravity plus matter, which is inspired by the previous SYK results, and turns out to provide a qualitatively similar picture in gravity. The parameter $k$ in gravity is analogous to $\kappa$ in SYK but we do not aim to establish a precise quantitative correspondence between the two.

\subsection{Equations of motion}

Let us now solve the equations of motion. The scalar field $\chi$ satisfies the massless wave equation
\be
\Box\chi = 0~.
\ee
Together with the condition \eqref{chiDirichlet} coming from the imaginary sources, this fixes the classical solution to be
\be\label{chiSol}
\chi  ={2i k \/\pi }\rho~.
\ee
We are finding here an imaginary solution for a real scalar field $\chi$ because of the complex sources. In Lorentzian signature, imaginary sources are unphysical as they lead to violations of the averaged null energy condition. Indeed, they can be used to construct traversable wormholes in AdS without a non-local coupling between the boundaries, in contradiction with the ``no-transmission principle'' \cite{Engelhardt:2015gla} (see \cite{Freivogel:2019lej} for a related discussion). In Euclidean signature, these imaginary sources do make sense. They can be defined by analytic continuation from real sources in the thermal partition function and can be used to study the statistical mechanics of a physical system. For example, imaginary chemical potentials have been useful in studying the phase diagram of gauge theories \cite{Roberge:1986mm, deForcrand:2002hgr}.

We now impose the equation of motion for the JT dilaton which is
\be\label{eqDilaton}
\n_\mu\n_\nu \Phi - g_\mn \Box\Phi + g_\mn\Phi +\ln T_\mn^\chi\rn=0 ~.
\ee
The stress tensor of $\chi$ decomposes into a classical and a quantum piece
\be
\ln T_\mn^\chi\rn=T_\mn^\r{class} +T_\mn^\r{quantum} ~.
\ee
The classical piece is obtained by evaluating $T_\mn$ on the solution \eqref{chiSol}, which leads to
\be
T_\mn^\r{class} = \p_\mu \chi \p_\nu\chi -{1\/2} g_\mn(\p\chi)^2 =  {2k^2\/\pi^2} \bpm 1 & 0 \\ 0 & -1 \epm~.
\ee
This generates negative energy in the bulk because the solution for $\chi$ is pure imaginary. What would normally be positive energy for real sources becomes negative energy for imaginary sources. 

In addition to this classical piece, there is also a quantum stress tensor due to the Casimir energy of $\chi$. Although this piece is negligible in the regime where we will compare with the previous field theory results, it becomes important at higher temperature, where it destabilizes the wormhole. The quantum stress tensor can be computed explicitly, using the Green function method and point splitting, as  detailed in Appendix \ref{QuantumTmn}. The result is
\be\label{TmnMatter}
T_\mn^\r{quantum}  = -{1\/24\pi} g_\mn -{\cE(b)\/\pi} \bpm 1 & 0 \\ 0 & -1 \epm~,
\ee
where we have introduced the quantity
\be
\cE(b) \equiv \sum_{n\geq 1} {1\ov 4\,\r{sinh}^2 ({nb\/2})} > 0 ~,
\ee
whose properties are given in the Appendix \ref{QuantumTmn}.

We can now solve for the JT dilaton using \eqref{eqDilaton}. The general solution is given by
\be
\Phi  =  \le( {2k^2\/\pi} - \cE(b)\ri)(1+\rho\,\r{tan}\,\rho) + {\eta\/2} \,\r{tan}\,\rho+ {1\/24\pi}~,
\ee
where $\eta$ is an arbitrary constant. The existence of the wormhole requires that $\Phi \to+\infty$ at both boundaries. Expanding this solution at the two boundaries gives
\bea
\Phi &\underset{\rho\to +{\pi\/2}}{\sim}& \le( {k^2\/\pi} - {\cE(b)\/2} +{\eta\/2}\ri)  {1\/{\pi\/2}-\rho} +\dots~,\\
\Phi &\underset{\rho\to -{\pi\/2}}{\sim} & \le( {k^2\/\pi} - { \cE(b)\/2} - {\eta\/2}\ri)  {1\/{\pi\/2}+\rho} +\dots~.
\eea
The position of the left and right boundaries are determined by imposing the boundary condition $\Phi = \bar\phi_r/\e$. We should also impose the boundary conditions for the metric
\be
ds^2 \underset{\rho\to {\pi\/2}}{\sim}{du_R^2\/\e^2},\qq ds^2 \underset{\rho\to -{\pi\/2}}{\sim}{du_L^2\/\e^2}~,
\ee
which tell us how $u_L$ and $u_R$ are related to $\tau$. The left and right temperatures are defined as their periods $u_L\sim u_L+\bar\phi_r\b_L, u_R\sim u_R+\bar\phi_r\b_R$ rescaled by $\bar\phi_r$ for convenience. This gives
\be
\b_L= {2b\/{2k^2\/\pi} -\,\cE(b)-\eta},\qq \b_R= {2 b\/{2k^2\/\pi}-\,\cE(b)+\eta}~.
\ee
We see that $b$ and $\eta$ are determined by the boundary conditions: they are fixed by the choice of temperatures on each side. The condition for the existence of the  wormhole is that $\b_L,\b_R\geq 0$. In the following, we will set the asymmetry parameter $\eta$ to zero so that we consider a situation where $\b_L=\b_R$.\footnote{The formulas for independent $T_L$ and $T_R$ can be obtained by replacing $T\ra{(T_L+T_R)/2}$.} 

The temperature $T=T_L=T_R$ is then
\be\label{Texplicit}
T = {k^2\/\pi b} - {\,\cE(b)\/2b}~.
\ee
At large $b$, this is positive since we have $\cE(b) \sim e^{-b}$ which becomes negligible. Hence, this gives a consistent wormhole solution. From the expression of the temperature, we see that the existence of the solution depends crucially on the pure imaginary sources. Real sources would change the sign of $k^2$ and prevent the solution to exist. Moreover, we see that the Casimir energy contributes negatively (since $\cE(b)>0$) indicating that it has a destabilizing effect which is described in more detailed below.

It's interesting to note that the operator $\cO$  acquires an imaginary  expectation value
\be\label{OWH}
\ln \cO_L\rn = {2ik\/\pi}~,\qq\ln \cO_R\rn = -{2ik\/\pi}~,
\ee
which can be read off from the solution for $\chi$ using the AdS/CFT dictionary. This is a consequence of the imaginary gradient in the  solution for $\chi$. We will see below that this expectation value is an order parameter for the phase transition.

It's interesting to note that the marginality of the operator $\cO$ is important for the wormhole solution to be possible. Indeed, an operator $\cO$ with $\D\neq 1$ doesn't lead to a consistent solution because it makes the dilaton grow too fast or too slow, see Appendix \ref{app:massive}. 

\newpage

\subsection{Thermodynamics}

We now compute the free energy of the wormhole solution. The partition function is
\bea\label{logZ}
\log Z\=   -b^2 T + {2 bk^2\/\pi}+ \log \Tr\,e^{- b L_0}~.
\eea
The first term comes from the classical contribution of the Schwarzian action that describes the two trumpet geometries \cite{Saad:2019lba} in conventions where $8\pi G_N =1$. The corresponding one-loop term proportional to $\log T$ is always negligible in our discussion and will be ignored. The second term is the on-shell action of the classical matter solution \eqref{chiSol}. The third term is the one-loop contribution from the matter, which can be derived as follows. The path integral of the scalar field $\chi$ on a cylinder of width $\pi$ and circumference $b$ gives the thermal partition function
\be
Z_\r{cylinder}^\chi = \Tr\,e^{-b \le( L_0-{1\/24}\ri)} = {1\/\eta(e^{-b})}~,
\ee
where $\eta(q)$ is the Dedekind eta function. We can perform a Weyl transformation to the double trumpet. It follows from the discussion in \cite{maldacena2018} that the effect of the Weyl anomaly is to shift the ground state energy by removing the $-{1\/24}$, so we end up with
\be
Z^\chi =  \Tr\,e^{-b L_0}={e^{-b/24}\/\eta(e^{-b})}~.
\ee 
This formula is also a consequence of the computation of the quantum stress tensor in Appendix \ref{QuantumTmn}.

The saddle point of \eqref{logZ} with respect to $b$ leads to the relation
\be\label{Tsaddle}
T = { k^2\/\pi b } - {\cE(b)\/2b}~,
\ee
where we used that $\cE(b)= {1\/24}+\p_b \log\eta(e^{-b})$ from the identities given in Appendix \ref{QuantumTmn}. Note that $\cE(b) = \ln L_0\rn_b$ is the thermal expectation value of $L_0$ at the inverse temperature $b$.\footnote{Without imaginary sources and for a general matter CFT, the expression \eqref{Tsaddle} becomes $T = -{\ln L_0\rn_b/(2b)}$ which is negative because $L_0$ is a positive operator. This shows that for any matter CFT, Casimir energy alone can never support the double trumpet geometry.} It is a nice consistency check that this expression of $T$ matches with \eqref{Texplicit} obtained from the explicit solution.

\pg{Large wormhole regime.} Let's first consider a regime where $b$ is large. In this regime, we have $\cE(b)\sim e^{-b}$ so the second term of \eqref{Tsaddle} becomes negligible as soon as $b$ becomes relatively large. We then have the saddle point $b_\ast = {k^2\/\pi T}$ which leads to the free energy
\be\label{FlargeWH}
F_\r{WH} = -T\log Z = -{k^4\/\pi^2}~,\qq \text{when }\quad \cE(b_\ast) \ll k^2
\ee
This formula is valid when $\cE\le({k^2\/\pi T}\ri)\ll k^2$ which holds at sufficiently low temperature since we have $\cE\le({k^2\/\pi T}\ri) \sim_{T\to0} e^{-k^2 /(\pi T)}$. We see that the wormhole has constant negative free energy at low temperature, as is expected from a gapped system. This is the regime where we expect that JT gravity and the SYK model share similar features.

\begin{figure}
	\centering
	\includegraphics[width=14cm]{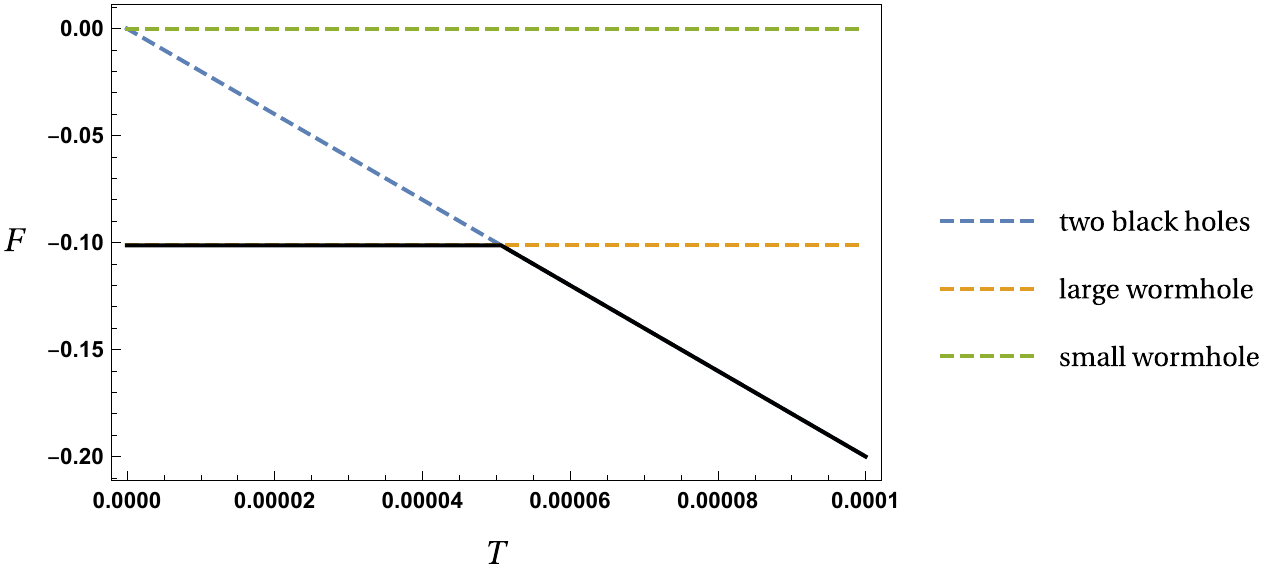}
	\caption{Plot of the free energy of the wormhole and the black holes at low temperature. The solid black line represents the free energy of the system. It corresponds with that of the wormhole, characterized by a constant negative free energy, at sufficiently low (high) temperatures. As temperature increases, we observe a first order transition separating the wormhole from the two black hole phase. This qualitatively matches the behavior seen in the SYK with complex couplings. Notice that there is also another solution at smaller $b$ whose free energy is always larger. We use $k=1$ and $S_0=10^3$.}\label{plotTransition}
\end{figure}

Our gravitational boundary conditions consist of two circles of length $\b=1/T$ and Dirichlet boundary conditions for the scalar field $\chi$ given in \eqref{chiDirichlet}. Besides the double trumpet solution, we can also have two disconnected black holes (\ie two hyperbolic disks). In a black hole, the source for the scalar field  doesn't contribute to the on-shell action, because the classical solution is just $\chi=\r{const}$. The one-loop contribution gives a contribution that depends on $\e$ and gets renormalized away. As a result, the free energy of the two black holes is the same as in pure JT gravity:
\be
F_\r{BH} = -T\log Z =- 2 S_0 T - 4\pi^2 T^2~.
\ee
The physical  free energy is then the minimum
\be
F =\min\le(F_\r{WH},F_\r{BH} \ri)~,
\ee
and there is a phase transition at the critical temperature 
\be
T_c \underset{k^2\ll S_0}{\sim} {k^4\/2\pi^2 S_0}~.
\ee
The classical solution $\chi=\r{const}$ in the black hole phase implies that the expectation value of the marginal operator vanishes:
\be
\ln \cO_L\rn = 0~,\qq\ln \cO_R\rn =0~,
\ee
This shows that the expectation value of $\cO$ is an order parameter for the transition from the two black holes to the wormhole. It is zero in the black hole phase, and becomes non-zero in the wormhole phase, where we have \eqref{OWH}. The non-zero expectation value at one boundary is a direct consequence of the source at the other boundary so this can be seen as a field theory incarnation of the geometric connection in the bulk.

\pg{Comparison with SYK.} The above computation gives the annealed free energy $-T \log\, \ln Z\rn$ where the ensemble average $\ln \,\cdot\,\rn $ is defined by the Euclidean path integral. To compare with the SYK setup, we should instead compute the quenched free energy $-T\ln \log Z\rn$. It was recently explained \cite{engelhardt2020} that the quenched free energy can be computed in gravity by including corrections from replica wormholes. These corrections are computed below in the next section \ref{sec:QA} and shown to give a negligible correction in the low temperature regime. This allows us to compare the gravitational free energy plotted in Fig.~\ref{plotTransition} to the SYK free energy shown in Fig.~\ref{fig:freeq}. The qualitative agreement of the free energies supports our proposal that the low-temperature phase observed in our SYK setup should be interpreted as a Euclidean wormhole.

\begin{figure}
	\centering
	\includegraphics[width=15cm]{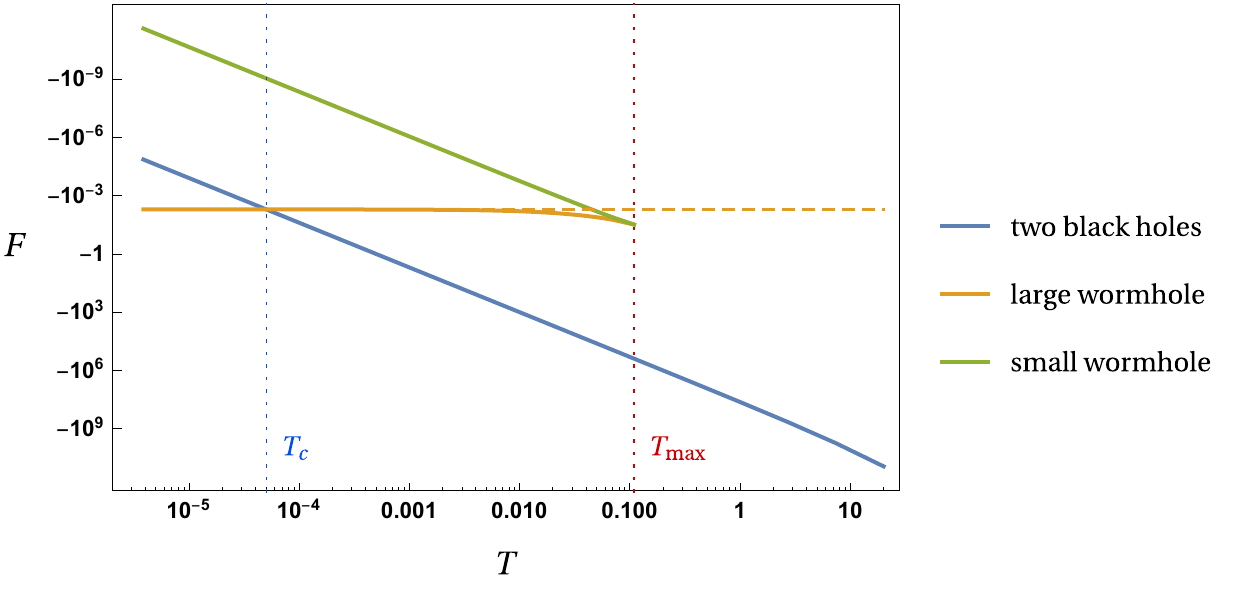}
	\caption{Log-log plot of the free energy of the different solutions. We plot the disconnected contribution of the two black holes (in blue). It dominates at high temperature until the phase transition at $T = T_c$ (dashed vertical blue line) after which the large wormhole solution (in orange) becomes favorable. There is also a small wormhole solution (in green) which has higher free energy. Both wormhole solutions only exist below a maximal temperature $T_\r{max}$ (dashed vertical red line) and we have $T_\r{max}>T_c$. The dashed orange line corresponds to the classical approximation where we don't include the Casimir energy of the matter, thus obtaining a wormhole at all temperatures. This shows that the Casimir energy  has a destabilizing effect which is responsible for the existence of both the maximal temperature and the  small wormhole solution. }\label{plotFtotal}
\end{figure}

\begin{figure}
	\centering
	\includegraphics[width=15cm]{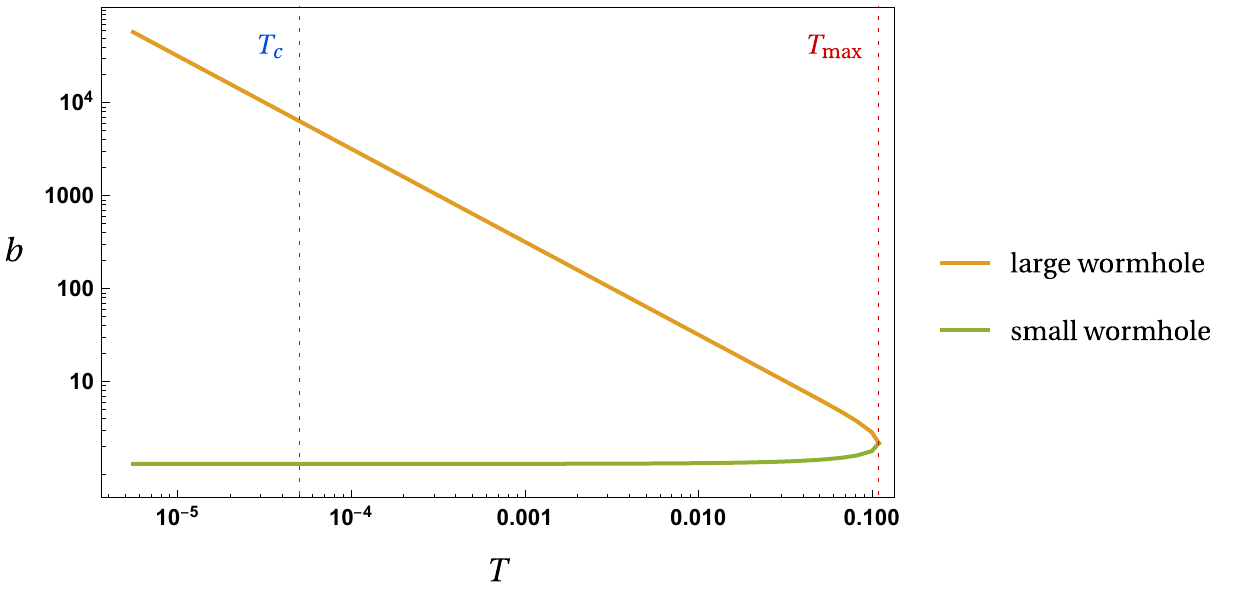}
	\caption{Log-log plot of the size $b$ of the wormhole as a function of the temperature. We use $k=1$ and $S_0=10^3$.}\label{plotSize}
\end{figure}

\pg{General analysis.} Let's now analyze the solution more generally, beyond the regimes where JT gravity is well approximated by SYK. Firstly, the wormhole solution only exists if $T\geq 0$. Since $\cE(b)$ is a decreasing function of $b$, we see that this implies that there is a minimal value of $b$  defined implicitly by
\be
\cE(b_\r{min})= {2 k^2\/\pi}~.
\ee
The formula \eqref{Tsaddle} also implies that there is a maximal value $T_\r{max}$ of the temperature beyond which the wormhole solution disappears. This value is attained when $\p_b T=0$ and the solution only exists when $0\leq T\leq T_\r{max}$. In this range, there are actually two values of $b$ that give the same $T$: the large wormhole that was discussed previously but also a small wormhole at a much smaller value of $b$. This is the result of a competition between the classical negative energy coming from the imaginary sources, and the positive quantum Casimir energy of the matter. It is here a classical effect that sustains the wormhole while a quantum effect destabilizes it.  It can be checked that the smaller wormhole has always higher free energy than the large one, so that it never dominates in the canonical ensemble.  The main thermodynamic features of the system are summarized in Fig.~\ref{plotFtotal} which describes the free energy of the different branches as a function of temperature and  Fig.~\ref{plotSize} that depicts the dependence of the wormhole size $b$ with temperature.

\subsection{Quenched versus annealed in gravity}\label{sec:QA}

\begin{figure}
	\centering
	\includegraphics[width=15cm]{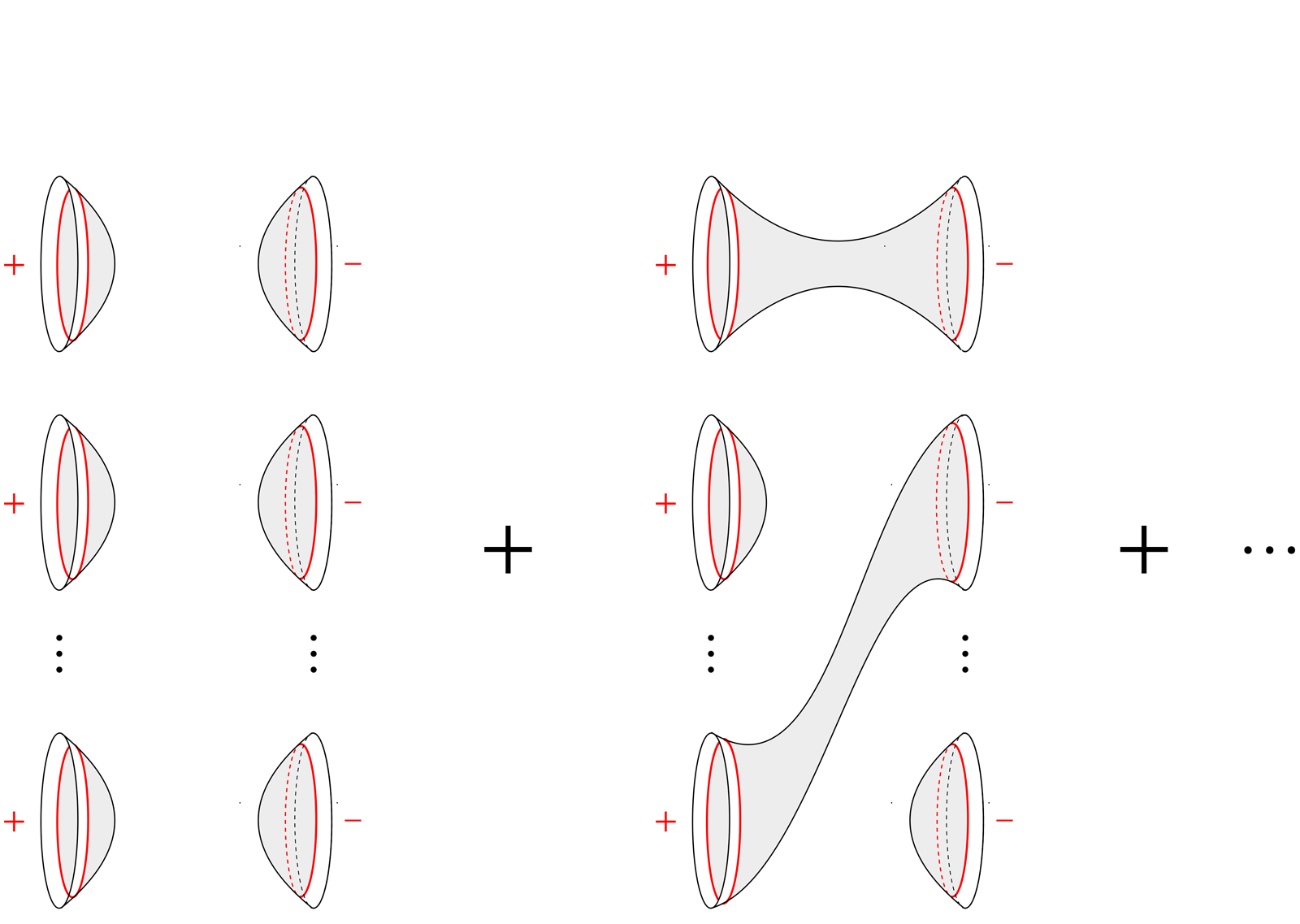}
	\caption{Computation of $\cP(B^m)$ in the replica trick for the free energy. }\label{fig:Freplicas}
\end{figure}

As pointed out in \cite{engelhardt2020}, the gravitational  free energy must be computed with a replica trick because we are interested in the quenched, rather than annealed, free energy. This means that the naive answer for the free energy (\ie the annealed one) might be incorrect in situations where quantum gravity effects are important. For a system with boundary  $B$, we should compute the gravitational path integral $\cP(B^m)$ whose boundary is $m$ copies of $B$, analytically continue in $m$ and use the formula
\be
\ln \log Z(B)\rn = \lim_{m\to0} {1\/m}(\cP(B^m)-1)~.
\ee
We take $B$ to be the system with two circle boundaries: one is labeled $+$ and has a source $+ik$ while the other is labeled $-$ and has the source $- ik$. The wormhole solution can only connect a $+$ circle to a $-$ circle. We take the temperature to be not too small so that taking $S_0$ to be large allows us to  ignore the higher genus topologies. A typical contribution to $\cP(B^m)$ is drawn in Fig.~\ref{fig:Freplicas}.

Denoting by $Z_1$ the black hole contribution and by $Z_2$ the wormhole contribution, we find
\be
\cP(B^m) = \sum_{k=0}^m k! \binom{m}{k} Z_2^k (|Z_1|^2)^{m-k}  = Z_2^m \exp\le({|Z_1|^2\/Z_2} \ri) \int_{|Z_1|^2/Z_2}^{+\infty} dt \, t^{m} e^{-t}~.
\ee
The effect of the replica trick is to introduce a permutation factor $k!$, which takes into account wormholes connecting boundaries belonging to different copies of $B$. Indeed, removing this factor leads to $\cP(B^m) = (|Z_1|^2 +Z_2)^m$ and gives the annealed result. The expression
\be
{1\/m}(\cP(B^m)-1)= \exp\le({|Z_1|^2\/Z_2} \ri) \int_{|Z_1|^2/Z_2}^{+\infty} dt \, \le( {1\/m}( Z_2^m t^m -1)\ri) e^{-t}~
\ee
makes it possible to take the $m\to0$ limit:
\be\label{mtozero}
\ln \log Z(B)\rn= \log Z_2 +e^{|Z_1|^2/Z_2} \int_{|Z_1|^2/Z_2}^{+\infty} dt \, \log \,t\, e^{-t} ~.
\ee
At large temperature $T>T_c$, there are only disconnected contributions and we obtain the free energy of two black holes. At small temperature $T<T_c$, we have $Z_2 \gg |Z_1|^2$ and \eqref{mtozero} gives
\be
\ln \log Z\rn= \log Z_2-\g~, \qq (T<T_c)~.
\ee
where we used $\int_0^{\infty} dt\,\log t\,e^{-t } = -\g$ with $\g\approx 0.577$ is the Euler gamma constant. This shows that the effect of the replica wormholes is to lower the entropy of the wormhole by $\g$. It gives the low temperature free energy 
\be
\ln F\rn=F_\r{WH} +\g T~, \qq (T<T_c)~,
\ee
where $F_\r{WH} =- {k^4/\pi^2} $ is the free energy \eqref{FlargeWH} of the wormhole solution. The second term represents the correction from the replica wormholes and can be neglected because  $T< T_c\ll k^4$. Hence, in our gravity setup, the difference between the quenched and annealed free energies becomes negligible at low temperatures. We can then compare the gravitational free energy plotted in Fig.~\ref{plotTransition} to the SYK free energy of Fig.~\ref{fig:freeq}. The qualitative agreement supports our proposal that the low-temperature phase observed in our SYK setup, which only arises after average over the complex SYK couplings, should be interpreted as a Euclidean wormhole.

\newpage

\section{Discussion}
One of the main motivation of this paper was to shed light on the holographic interpretation of Euclidean wormholes.  We have
 shown how thermodynamic properties consistent with a wormhole can arise from averaging over complex couplings in the SYK model, and we have proposed a gravity interpretation as  a Euclidean wormhole solution of JT gravity plus matter sharing  similar properties.

We have studied the free energy of two copies of a system with complex conjugated Hamiltonians. This is also equal to the real part of the free energy of a single copy:
\be
2 \, \r{Re}\, \ln F\rn      =  -T \ln \log Z \bar{Z}\rn ~.
\ee
From the point of view of a single system, our wormhole can thus be seen as ``real part wormhole'' which connects the system to its complex conjugate in the gravitational computation of $\r{Re}\, F$. The gravitational computation of the quenched free energy also involves replica wormholes which were computed in section \ref{sec:QA} and shown to be negligible in our setup.

The mechanism that makes the Euclidean wormhole possible is the introduction of imaginary sources for a marginal operator, which provides the negative energy necessary to sustain the wormhole. We emphasize that adding imaginary sources is perfectly consistent in Euclidean signature; it is akin to probing a system at imaginary chemical potential which, for example, has been useful  to study QCD  \cite{Roberge:1986mm, deForcrand:2002hgr}. We have identified a low temperature phase, where the wormhole dominates, and in which the marginal operator condenses. The non-zero expectation value of this operator is linked to the existence of the wormhole, since it vanishes in the black hole phase and becomes non-zero in the wormhole because of the source on the other boundary. This can be seen as a field theory diagnosis of the geometric connection in the bulk.

Our wormhole is different from the eternal traversable wormhole of \cite{maldacena2018}, which makes sense as a Lorentzian geometry, and requires an explicit coupling between the boundaries. It's also possible to introduce such an explicit coupling in our setup, in addition to the imaginary sources, and it would be interesting to study the resulting solutions.

In this paper, we have seen that imaginary sources  can be used to support Euclidean wormhole solutions. As the comparison with SYK suggests, these sources might be related to some averaging procedure. It would be interesting to see whether other Euclidean wormholes can be constructed using this idea. Our solution  extends to JT gravity with an additional gauge symmetry \cite{Iliesiu:2019lfc,Kapec:2019ecr} or including the gravitational $\r{U}(1)$ symmetry described in \cite{Godet:2020xpk}. It should be possible to study multi-boundary solutions for which a dual SYK setup might exist. Euclidean wormholes have also found recent applications in AdS$_3$ holography \cite{Maxfield:2020ale,Maloney:2020nni,Afkhami-Jeddi:2020ezh, Cotler:2020ugk,Belin:2020hea} and it would be interesting to explore solutions in higher dimensions \cite{Cotler:2020lxj}.

 Euclidean wormholes have played an important role in the study of eigenvalue statistics in pure JT gravity and in its field theory realization as a random matrix model \cite{Saad:2018bqo, Saad:2019lba, Saad:2019pqd,garciagarcia2019}.  The conclusion of these works is that JT gravity is quantum chaotic. Level statistics of quantum chaotic systems are described by random matrix theory. For the spectral form factor, a useful observable in spectral analysis, a signature of quantum chaos is the presence of a ramp for sufficiently long times that saturates at the Heisenberg time. In pure JT gravity, the spectral form factor is related to a double trumpet geometry which is not a solution of the equations of motion. An explicit evaluation of the path integral gives the contribution
 \be\label{JTDT}
\le\ln Z(\b_L)Z(\b_R)\ri\rn_\r{conn}^\r{JT} = {1\/2\pi}{\sqrt{\b_L\b_R}\/\b_L+\b_R} +\text{higher genus}~,
\ee
which is actually the universal answer for a double-scaled random matrix integral. Replacing  $\b_L = \b+it$ and $\b_R=\b-it$ leads to a contribution that grows linearly with $t$ for $t\gg\b$ and accounts for the ramp in the spectral form factor. The same computation does not work for JT gravity plus matter as the integral over $b$ has a divergence at small $b$, seen here from the fact that $\cE(b)\sim_{b\to0} {\pi^2 \/ 6 b^2}$. If we add imaginary sources, we can get around this issue by using our wormhole solution as a saddle point in this path integral. This suggests that Euclidean wormhole solutions, of the type constructed here, might be useful in studying the eigenvalue statistics of gravitational theories for which we cannot perform the full path integral. We leave this interesting question and the comparison with SYK level statistics to a future work.

 Euclidean wormholes have also been important in understanding better the fine-grained entropy of evaporating black holes, where replica wormholes were crucial in obtaining an answer consistent with unitarity \cite{penington2020,almheiri2020}. It has also been argued that a gravitational replica trick is   already needed for the computation of the free energy \cite{engelhardt2020}. We have implemented this replica trick in section \ref{sec:QA} to compute the free energy of our gravity setup and shown that the additional wormholes only give a negligible correction. It would be interesting to see whether wormhole solutions similar to the one described in this paper could be used as saddle points in the replica computation of the free energy in situations where they give large corrections, for example at very low temperatures.

On the field theory side, a non-Hermitian Hamiltonian is superficially related to a loss of probability conservation and a non-unitary evolution as most eigenvalues are complex and therefore have a finite lifetime. From the point of view of one of the  two systems, 
this could be interpreted as the observation of particles coming in and coming out which is typical of an open quantum system. Although this is an appealing picture, it cannot really explain why a gap is observed. Ensemble average is a key ingredient for the formation of the gap at low temperatures so just a spectrum with imaginary eigenvalues is not enough to reproduce the observed phenomenology. As we mentioned earlier, another outstanding feature is that for sufficiently high temperature the non-Hermitian effects related to a complex spectrum becomes irrelevant. This suggests that the addition of an imaginary part in the SYK couplings, and the subsequent ensemble average, is just an effective way to describe quantum tunneling between two Hermitian SYK's, and produces a gap between the ground state and the first excited state as in a double well potential. Typically, this is qualitatively modeled by an explicit coupling between the two SYK's, as in the case the eternal traversable wormhole. However, we are interested in understanding better the issue of factorization, so we want to study configurations with an explicit coupling between the two systems.

Assuming that complex couplings and ensemble average are necessary ingredients to model tunneling without an explicit coupling, it would be interesting to determine the conditions on the field theory Hamiltonian that lead to this type of wormhole behavior. It seems that some form of randomness is required as wormhole-like features are only seen after an ensemble average. However, it is unclear whether infinite-range and strong interactions, as in the SYK model studied here, are also a necessary requirement. 

\acknowledgements
 AMG was partially supported by the National Natural Science Foundation of China (NSFC) (Grant number 11874259) and also acknowledges financial support from a Shanghai talent program. AMG acknowledges illuminating correspondence with Zhenbin Yang, Jac Verbaarschot, Dario Rosa and Yiyang Jia. VG acknowledges useful conversations with Charles Marteau.

\appendix

\section{Quantum stress tensor in the double trumpet}\label{QuantumTmn}

In this section, we compute the quantum stress tensor of a massless scalar in the double trumpet geometry. We consider a free boson described by the action
\be
S ={1\/2} \int d^2 x \sqrt{g}\,  (\p\chi)^2~,
\ee
and whose stress tensor is given by
\be
T_\mn =\p_\mu \chi\p_\nu\chi - {1\/2} g_\mn (\p\chi)^2~.
\ee
To compute its expectation value, we use the point-splitting method so that
\be\label{pointSplitT}
\ln T_\mn (x) \rn = \lim_{x'\to x} \le( \p_\mu \p_\nu' G(x,x') -{1\/2} g^\mn \p_\mu \p_\nu' G(x,x')  \ri)~,
\ee
where $G(x,x')$ is the Green function. See \cite{Milton:2001yy,Freivogel:2019lej} for examples of application of this method in related contexts. We will first perform the computation on a cylinder described by the coordinates
\be
ds^2_\r{cyl} = d\rho^2+d\t^2~,\qq -{\pi\/2} \leq \rho\leq {\pi\/2}~,\qq \t\sim \t+b~.
\ee
We will then use a Weyl transformation to the double trumpet. The Green function can be defined as the solution of the equation
\be\label{boxGeq}
\Box_x G(x,x') = -\d(x-x')~.
\ee
Using the following mode decomposition
\be
G(\rho,\t;\rho',\t') =\sum_{\substack{ m,m'\in\Z}}G(\rho,m;\rho',m')\,  e^{2i\pi (m\t+m'\t')/b}~,
\ee
the equation \eqref{boxGeq} becomes
\be\label{eqGmm}
\le(  \p_\rho^2 -{4\pi^2 m^2\/b^2}\ri)G(\rho,m;\rho',m')    = -{1\/b}\d(\rho-\rho') \d_{m+m'}~,
\ee
with the boundary conditions
\be\label{bdyGmm}
G(\rho,m;\rho',m')|_{\rho=\pm {\pi\/2}}= G(\rho,m;\rho',m')|_{\rho'=\pm {\pi\/2}} = 0~.
\ee
This can be solved using the discontinuity method, as explained in   \cite{Milton:2001yy} for the standard Casimir effect between two plates. At the end, we find
\be
G(\rho,\t;\rho',\t') =-\sum_{m\in\Z}{\r{sinh}({2\pi m\/b}(\rho_+-\tfrac\pi2))\,\r{sinh}({2\pi m\/b}(\rho_-+\tfrac\pi2))\/2\pi m \,\r{sinh}({2\pi^2m\/b})}  e^{2i\pi m(\t-\t')/b}~,
\ee
which allows us to obtain
\be
\ln T_{\rho\rho}\rn =- \sum_{m\in \Z}{ \pi m\/b^2\r{tanh}\le({2\pi^2 m\/b} \ri)}~,\qq \ln T_{\t\t}\rn = \sum_{m\in \Z}{ \pi m\/b^2\r{tanh}\le({2\pi^2 m\/b} \ri)}~.
\ee
The sum over $m$ can be regularized using the Poisson resummation formula, which allows us to write
\be\label{Sfirst}
S(x) \equiv\sum_{m\in \Z } {m\ov \r{tanh}( m x)} = -{\pi^2\/2x^2} \sum_{n\in \Z} {1\/\r{sinh}^2({\pi^2 n \/x})} ~.
\ee
We regulate the divergence at $n=0$ by imposing that in the limit $b\to+\infty$, we recover the known result for the strip. The final formula is
\be\label{TrhorhoReg}
\ln T_{\t\t}\rn = -\ln T_{\rho\rho} \rn ={c\/24\pi} -{c\/\pi}\cE(b)~,
\ee
where have multiplied by an overall $c$ and defined 
\be
\cE(b) \equiv \sum_{n\geq 1} {1\ov 4\,\r{sinh}^2 ({nb\/2})}~.
\ee
We can now compute the stress tensor in the double trumpet using the formula
\be
\ln T_\mn\rn = \ln T_\mn\rn_\r{cyl} + \ln T_\mn\rn_\r{anomaly}~,
\ee
where the formula for the anomalous stress tensor is given in \cite{Brown:1977sj}. This finally gives in the $(\tau,\rho)$ coordinates
\be
\ln T_\mn\rn  = -{1\/24\pi} g_\mn -{\cE(b)\/\pi} \bpm 1 & 0 \\ 0 & -1 \epm~.
\ee

\pg{Some properties of the function $\cE(b)$.}

We give some properties of the function $\cE(b)$ which appeared in the above computation. Firstly, we have  $\cE(b)>0$ and $\cE(b)$ decreases with the following asymptotics:
\be
\cE(b)\underset{b\to0}{\sim} {\pi^2 \/6 b^2}~,\qq \cE(b) \underset{b\to+\infty}{\sim} e^{-b} + O(e^{-2b})~.
\ee
 After defining $q=e^{-b}$, we can write $\cE(b)$ as
\bea
\cE(b) =\sum_{n\geq 1} {q^{n}\/(1-q^{n})^2} \= \sum_{n\geq 1}\sum_{m\geq1} m  q^{nm}  = \sum_{m\geq 1} { mq^{m}\/1-q^{m}}={1\/24}(1-P(q)) ~,
\eea
in terms of the function $P$ studied by Ramanujan \cite{berndt1985ramanujan}, which is related to the Dedekind eta function 
\be
P(q) = 24 q {d\ov dq}\log  \eta(q)~,\qq \eta(q) \equiv q^{1/24}\prod_{n\geq 1}(1-q^n)~.
\ee
There is no simple analytic expression for $\cE(b)$, but it can be expressed in a complicated way in terms of elliptic integrals \cite{Berndt1998}. 

\section{Massive bulk fields}\label{app:massive}

We have constructed a wormhole solution using a massless scalar field in the bulk, corresponding to an operator $\cO$ with conformal dimension $\D=1$. In this Appendix, we consider massive fields and explain why $\D=1$ is the only value that gives a consistent solution. 

The main point is that we need to satisfy the two boundary conditions of JT gravity:
\be
\Phi = {\bar\phi_r\/\e},\qq ds^2 = {du^2\/\e^2}~.
\ee
Satisfying both in the double trumpet \eqref{doubleTrumpet} requires that at leading order
\be\label{phiasymp}
\Phi \underset{\rho\to {\pi/2}}{\sim} {C\/{\pi\/2}-\rho}~,
\ee
where $C$ is some constant. Now, take $\chi$ to be a general massive scalar field satisfying the equation
\be
(\Box-m^2)\chi = 0~,\qq m^2=\D(\D-1)~.
\ee
The pure imaginary sources give the condition
\bea\label{bdycondchi}
\chi \underset{\rho\to{\pi\/2}}\sim  i k \le({\pi\/2}-\rho \ri)^{1-\D} ,\qq \chi \underset{\rho\to-{\pi\/2}}\sim -ik \le({\pi\/2}+\rho \ri)^{1-\D} ~.
\eea
This fixes the classical solution for $\chi$ which can be written explicitly in terms of hypergeometric functions.

We will focus on the low-temperature regime where the Casimir energy of $\chi$ can be ignored, which is the relevant regime for comparison with SYK. There, the equation of motion for the JT dilaton is
\be
\n_\mu\n_\nu \Phi - g_\mn \Box\Phi + g_\mn\Phi + T_\mn^\chi=0 ~,\qq T_\mn^\chi=\p_\mu \chi \p_\nu\chi -{1\/2} g_\mn \le((\p\chi)^2+m^2\chi^2\ri)~.
\ee
Taking the trace of this equation gives
\be
(\Box-2)\Phi = - \D(\D-1)\chi^2~.
\ee
From \eqref{phiasymp}, we can verify
\be
\lim_{\rho\to {\pi\/2}}(\Box-2)\Phi =\text{finite constant}~,
\ee
where the constant depends on subleading terms in $\Phi$. This is only compatible with $\D=1$ because otherwise $\chi^2\sim \le({\pi\/2}-\rho \ri)^{-2(\D-1)}$ diverges for $\rho\to {\pi\/2}$. This shows that a wormhole solution can be obtained in this way  only for $\D=1$.

\bibliography{librarynh}
 
\end{document}